\documentclass[pra,twocolumn,epsf]{revtex4}

\input epsf

\newcommand{\be}[1]{\begin{equation}}
\newcommand{\ee}[1]{\end{equation}}
\begin{document}
\title{The dynamics of an ion chain in a harmonic potential} 
\author{Giovanna Morigi$^1$ and Shmuel Fishman$^2$} 
\address{$^1$ Department of Quantum Physics, University of Ulm,
  Albert-Einstein-Allee 11, 
D-89069 Ulm, Germany\\ 
$^2$ Department of Physics, Technion, 32000 Haifa, Israel} 

\date{\today}

\begin{abstract}
Cold ions in anisotropic harmonic potentials can form ion
chains of various sizes. Here, the density of ions is not uniform, thus
the eigenmodes are not phononic-like waves. 
We 
study chains of $N\gg 1$ ions and evaluate analytically the long wavelength
modes and the density of states in the short wavelength limit. These results
reproduce with good approximation
the dynamics of chains consisting of dozens of ions. Moreover, they 
allow to determine the critical transverse frequency required for the
stability of the linear structure, which is found in agreement with results
obtained by different theoretical methods [D.H.E.\ Dubin, Phys.\ Rev.\ Lett.\ {\bf 71}, 2753
(1993)] and by numerical simulations [J.P. Schiffer, Phys.\ Rev.\ Lett.\ {\bf 70}, 818 (1993)].
We introduce and explore the thermodynamic limit for the ion chain. The
thermodynamic functions are found to exhibit deviations from
extensivity.  
\end{abstract}
\maketitle
\section{Introduction}
Coulomb crystals are organized structures of charged particles, which interact
through the Coulomb repulsion and organize in regular patterns 
at sufficiently low temperatures in presence of a confining
potential~\cite{DubinRMP}. These 
potentials are realized by means of Paul or of Penning
traps~\cite{Ghosh}, and their geometry determines the crystal's structure. 
Several remarkable experiments have reported crystallization of ion gases in
Paul traps~\cite{Bluemel1988,Wineland1987,Raizen,MPQ,Drewsen}, 
Penning traps~\cite{Gilbert1988} and ion--storage rings~\cite{Habs}. The
Bragg-scattering in three--dimensional structures was studied providing
information about the internal structure of the crystal~\cite{Tan95,Itano98}. 
These crystals represent a kind of rarefied condensed
matter, the inter-particle
distance being of the order of several micrometers, allowing to study the
structure by means of optical radiation. Variation of the potential permits to
control the crystal shape as well as the number of ions, thus offering
the unique opportunity of studying the 
transition from few particles to mesoscopic systems.
Besides, these structures have been object of growing interests
as they provide promising applications for spectroscopy~\cite{Kienle,Here1}, 
frequency standards~\cite{Here2}, study and control of chemical
reactions~\cite{DrewsenChemistry}, and quantum information
processors~\cite{Cirac95,Steane,QLogic:2,QLogic:3}.    

In this work we investigate the dynamics of Coulomb chains.
These are one--dimensional structures, obtained by means of strong transverse
confinement and that usually consist of dozens of ions localized 
along the trap axis~\cite{Raizen,MPQ}. 
They represent a peculiar
crystallized structure: In fact, due to the axial potential 
the equilibrium charge distribution is
not uniform~\cite{Dubin97,Schiffer03}. 
This is in contrast to the three--dimensional case, where the density
of charges in a harmonic potential is uniform and, therefore, where 
the eigenmodes are phononic-like waves. In the Coulomb chain 
the non-uniformity of the density of ions combined with the
long-range interaction result in excitations that are fundamentally different
from the phonons in solids and lead to interesting thermodynamic properties. 
The exploration of these excitations and of the
chain thermodynamics is the subject of the present paper.

Our starting point is the ions equilibrium configuration evaluated
in~\cite{Dubin97}. We investigate the dynamics for small oscillations, when
the harmonic approximation is valid,
in the limit of large number of ions. We derive analytically
the eigenfrequencies and the
corresponding eigenmodes for
the long wavelength excitations. These are compared
with numerical results and good agreement is
found. From the resulting dispersion relation the density of states of the
long wavelength eigenmodes is determined. An analytic form of the density
of states is also found for the short wavelength excitations. 
This result allows to evaluate the critical
aspect ratio between the frequencies of the transverse and the axial confining
potential, that determines the stability of the chain. The value we find
agrees with numerical results~\cite{Schiffer93}, which 
have been experimentally verified for chains of few ions~\cite{LosAlamos00}. 
In particular, it is in agreement with the analytical estimate 
in~\cite{Dubin93}, which was obtained under different requirements. \\
\indent 
Using these results we discuss the statistical mechanics of the chain, and
derive some thermodynamic quantities in a specific thermodynamic limit, that
is defined here by keeping constant the density of ions in the chain center,
as the number of ions tends to infinity and the
axial frequency to zero, analogously to the 
definition for cold neutral gases in
traps~\cite{Stringari}. Non-extensive thermodynamic
properties are found. We compare the thermodynamic functions
with the ones of a chain of finite
number of ions that are obtained numerically, 
and find reasonable agreement.

This work is organized as follows. In section II we introduce the basic
equations and discuss the fundamental properties. In section III the spectrum
of excitations is studied. In section IV we investigate the statistical
mechanics of the system. Section V presents the conclusions and the outlook. In
the appendices, several details of the calculations of section III are
reported.  

\section{A string of charges in a harmonic potential}

The Hamiltonian describing the dynamics of a chain of $N$ ions of mass $m$ and
charge $Q$, which are confined by a harmonic potential, is given by
\begin{equation} \label{H:0}
H=\sum_{j=1}^N\frac{{\bf p_j}^2}{2m}+
V({\bf r_{1}},\ldots,{\bf r_{N}}) 
\end{equation}
where ${\bf r_j}=(x_j,y_j,z_j)$, ${\bf p_j}$ are the positions and conjugate
momenta ($j=1,\ldots, N$). 
The term $V$ accounts for the oscillator's potential and the 
Coulomb repulsion, 
\begin{eqnarray}
V &=&\frac{1}{2}\sum_jm\left(\nu^2x_j^2+\nu_t^2(y_j^2+z_j^2)\right)\\
&+&\frac{1}{2}\sum_{j=1}^N\sum_{j\neq i}\frac{Q^2}{\sqrt{(x_i-x_j)^2+
(y_i-y_j)^2+(z_i-z_j)^2}}\nonumber
\end{eqnarray}
where the harmonic oscillator has
rotational symmetry around the $x$-axis with axial and transverse 
frequencies $\nu$, $\nu_t$, respectively. 

For sufficiently low kinetic energy crystallization occurs. The temperature, at
which the gas is crystallized, corresponds to large plasma parameters
$\Gamma=Q^2/a_{\rm WS} k_BT\gg 1$. Here, $a_{\rm WS}$ is the Wigner-Seitz
radius, that is a function of the mean density $n$ and is defined as
$a_{\rm WS}=(3/4\pi n)^{1/3}$~\cite{DubinRMP}. In this regime the ions
are localized at the classical equilibrium positions ${\bf r_j^{(0)}}$, 
that satisfy the equations $\partial V/\partial {\bf r_j}|_{\bf r_j^{(0)}}=0$,
and such that the potential energy is minimal. When the harmonic potential
is sufficiently asymmetric, i.e.\ for  $\nu\ll \nu_t$, 
the ions equilibrium positions are confined to the trap
axis~\cite{Schiffer93}, namely ${\bf r_j^{(0)}}=(x_j^{(0)},0,0)$, 
and satisfy the equation describing the equilibrium of the forces,
\begin{eqnarray}
\label{Eq:Discrete} m\nu^2x_i^{(0)} =
-\sum_{j>i}\frac{Q^2}{(x_j^{(0)}-x_i^{(0)})^2}+ 
\sum_{j<i}\frac{Q^2}{(x_i^{(0)}-x_j^{(0)})^2}
\end{eqnarray} 
where the numbering convention is $x_i>x_j$ for $i>j$.
The stability of these points with respect to the transverse vibrations
depends on the number of ions $N$ and on the 
ratio $\nu_t/\nu$, and it is discussed in Section~\ref{Sec:Tr}. In this
section, we assume that the configuration is stable and approximate the
potential by its second order Taylor expansion around 
the points ${\bf r_j^{(0)}}$. We denote 
by $q_i=x_i-x_i^{(0)}$ the displacements in the $\hat{x}$-direction, and 
approximate the Hamiltonian (\ref{H:0}) as $$H\approx V_0+H_{\rm har},$$ where
$V_0=V({\bf r_{1}^{(0)}},\ldots,{\bf r_N^{(0)}})$ is the classical minimum
energy, while $H_{\rm har}$ 
describes the (classical) harmonic oscillations around the equilibrium
points~\cite{James,Kielpinski,EPJD},  
\begin{eqnarray}
\nonumber 
H_{\rm har}
&=&\sum_{j=1}^N\frac{{\bf p_j}^2}{2m}+\frac{1}{2}\sum_jm\nu^2q_j^2+
\frac{1}{2}\sum_jm\nu_t^2\left(y_j^2+z_j^2\right)\\ 
&+&\frac{1}{4}\sum_{i}\sum_{j\neq i}K_{i,j}(q_i-q_j)^2\nonumber\\
&-& \frac{1}{8}\sum_{i}\sum_{j\neq
i}K_{i,j}\left((y_i-y_j)^2+(z_i-z_j)^2\right)
\label{H:1}
\end{eqnarray} 
and the coefficients $K_{i,j}=\partial^2 V/\partial x_j^2|_{\{x_l^{0}\}}$ are 
positive and take the form
\begin{equation} 
\label{Kij}
K_{i,j}=\frac{2Q^2}{|x_i^{(0)}-x_j^{(0)}|^3}. 
\end{equation} 
Equation~(\ref{H:1}) shows that the axial motion is decoupled
from the transverse motion in the harmonic expansion. The corresponding 
equations of motion are
\begin{eqnarray} 
\label{Eq:ax}
&&\ddot{q}_i=-\nu^2q_i-\sum_{j\neq i}\frac{K_{i,j}}{m}(q_i-q_j)\\
&&\ddot{y}_i=-\nu_t^2y_i+\frac{1}{2}\sum_{j\neq i}\frac{K_{i,j}}{m}(y_i-y_j)
\label{Eq:y}\\ 
&&\ddot{z}_i=-\nu_t^2z_i+\frac{1}{2}\sum_{j\neq i}\frac{K_{i,j}}{m}(z_i-z_j) 
\label{Eq:z}
\end{eqnarray}  
and describe a system of coupled oscillators, with long range interaction and 
position-dependent coupling strength. In this paper eigenmodes will be 
calculated. For this we assume $q_i(t)=\int 
{\rm e}^{{\rm i}\omega t}\tilde{q}_i(\omega){d}\omega/2\pi $. 
To simplify notations we replace
$\tilde{q}_i(\omega)$ by $q_i$. This results in equations for the eigenmodes
of frequency $\omega$ that are similar to (\ref{Eq:ax}), but with 
$\ddot{q}_i$ replaced by $-\omega^2q_i$. The same replacement will be 
performed for $y_i(t)$ and $z_i(t)$.

It can be easily verified that the center-of-mass motion is an eigenmode of 
the secular equations~(\ref{Eq:ax}-\ref{Eq:z}). 
The axial center-of-mass mode is 
$q_1=\ldots=q_N$ at the characteristic frequency is $\nu$, while the
transverse center-of-mass modes are $y_1=\ldots=y_N$ and $z_1=\ldots=z_N$ at
frequency $\nu_t$. 
We remark that the axial and the transverse coupling terms appearing in
Eqs.~(\ref{Eq:ax}-\ref{Eq:z}) have 
opposite signs. Due to this property, in the axial 
direction the collective excitations are of higher frequencies than the 
center-of-mass frequency $\nu$, while in the transverse plane the 
collective excitations are of lower frequencies than $\nu_t$.

\subsection{Properties and symmetries}

Hamiltonian~(\ref{H:1}) is not translationally invariant, and this is a
consequence of the non-uniformity of the ions equilibrium distribution, 
due to the harmonic force appearing in Eq.~(\ref{Eq:Discrete}). 
The Hamiltonian~(\ref{H:1}) is however invariant under
reflection with respect to the center of the trap, which coincides with the
origin of the axes. In particular, $$x_i^{(0)}=-x_{-i}^{(0)}$$ where
$i=1,\ldots, N'$ (here, $N'=N/2$ for even $N$, while $N'=(N-1)/2$ for odd $N$).
Hence, the normal modes of the chain are symmetric (even) or 
antisymmetric (odd) under reflection with respect to the
center~\cite{Kielpinski,EPJD}, such that   
\begin{equation}
\label{Symmetry}
w_i^{(n)}=\pm w_{-i}^{(n)}
\end{equation}
with $w_i^{(n)}=q_i^{(n)},y_i^{(n)},z_i^{(n)}$, and $n$ labels the mode.
Some general properties can be inferred from this simple consideration. 
For instance, the even modes of the axial motion are characterized by 
constant length $L$ of the chain, since $q_{N'}^{(n)}=q_{-N'}^{(n)}$. 
For the odd modes, on the other 
hand, the center of mass of the chain, which coincides with the chain center,
does not move. Clearly, the center-of-mass mode, which we
denote by $w_i^{(1)}$, is an even mode characterized by equal displacements at
the positions 
$x_i^{(0)}$ of the chain. This property and the orthogonality 
between the normal modes lead to the relation
$$\sum_j w_j^{(n)}=0$$
for all normal modes with $n>1$.

It is remarkable that also the lowest axial odd mode (stretch mode)
and its frequency can be exactly determined. In fact, taking $q_i^{(2)}\propto 
x_i^{(0)}$ and substituting into (\ref{Eq:ax}) one finds
\begin{eqnarray}
\nonumber
(\omega^2-\nu^2)q_i^{(2)}
&=&-\sum_{j>i}\frac{2Q^2/m}{(x_j^{(0)}-x_i^{(0)})^2}+
\sum_{j<i}\frac{2Q^2/m}{(x_i^{(0)}-x_j^{(0)})^2} \nonumber\\
&=&2\nu^2q_i^{(2)}
\end{eqnarray}
where we have used relation~(\ref{Eq:Discrete}). 
Therefore, the frequency of the axial
stretch mode $q_i^{(2)}$ is $\sqrt{3}\nu$ and its value is independent of the
number of ions $N$ of the chain. This property has been first demonstrated
in~\cite{Dubin&Schiffer}. It has also been observed 
by numerical evaluation of the normal modes of chains up to 10
ions~\cite{Steane,James}. 
Analogously, the transverse stretch mode, which is the 
highest odd transverse excitation, satisfies
$y_i^{(2)},z_i^{(2)}\propto x_i^{(0)}$ with
eigenfrequency $\sqrt{\nu_t^2-\nu^2}$, which is also independent of $N$. 

We remark that the invariance under reflection imposes different boundary 
conditions than the ones that are usually chosen for a crystal with uniform ion
distribution. In a crystal that is translationally
invariant even and odd modes are degenerate and one may
choose periodic boundary conditions~\cite{Ashcroft}.
In presence of an external potential with central symmetry 
this invariance is broken, apart for the mirror symmetry with respect to 
the center. Hence,  at the edges the eigenmodes fulfill the relation
$w_{-N'}=\pm w_{N'}$, where the sign is determined by the parity. 

\section{The secular problem}

The systematic derivation of an analytic solution of  
Eqs.~(\ref{Eq:ax}-\ref{Eq:z}) is a challenging problem, since
it requires to take systematically into account the position dependent
coupling constant and the long range interaction.
Nevertheless, in the limit of large number of ions $N\gg 1$
we can make some simplifying assumptions. In this limit, in fact, 
the interparticle spacing 
$a_L(x_i)=x_{i+1}^{(0)}-x_i^{(0)}$ is a smooth function of the position, 
and it is inversely proportional to the density 
of ions per unit length~\cite{Dubin97},
\begin{equation} 
\label{Abstand} 
n_L(x_i)=1/a_L(x_i). 
\end{equation} 
The density of charges for unit length can be evaluated by applying the Gauss 
theorem to a continuous distribution of charges, that is assumed to be
uniformly distributed in an elongated ellypsoid.
The resulting one dimensional density is~\cite{Dubin93}
\begin{equation} 
\label{Gauss}
n_L(x)=\frac{3}{4}\frac{N}{L}\left(1-\frac{x^2}{L^2}\right) 
\end{equation} 
which is defined for $|x|\le L$, where $2L$ is the length of the crystal at 
equilibrium. The density (\ref{Gauss}) gives a good estimate of the charge
distribution in the center of the chain for $N$ sufficiently 
large~\cite{Dubin97}. The length $L$ is evaluated
by minimizing the energy of the crystal, and at leading
order in $N$ fulfills the relation~\cite{Dubin97}
\begin{eqnarray} 
\label{Length}
L(N)^3=3\left(\frac{Q^2}{m\nu^2}\right)N\log N.
\end{eqnarray} 
In the following, we use these quantities to
derive an approximate solution for the long- and
short-wavelength modes in the limit of $N\gg 1$ ions.
Furthermore, we compare the results of the derivation with the 
numerical calculations, obtained by solving the eigenvalue
equations~(\ref{Eq:ax}-\ref{Eq:z}) for a finite number of ions.

\subsection{The eigenmodes in the long wavelength limit}
\label{Sec:Jac}

We use the ansatz $q_i(x,t)={\rm e}^{{\rm i}\omega t}\tilde{q}_i(x)$ in
Eqs.~(\ref{Eq:ax}) and 
define the rescaled positions $\xi_i^{(0)}=x_i^{(0)}/L$ and the rescaled 
interparticle distances
$a(\xi_i)=a_L(x_i)/L$. With these definitions, denoting for simplicity of
notation $\tilde{q}_i\to q_i$, Eqs.\ (\ref{Eq:ax}) take
the form
\begin{equation} 
\label{Ax:Rescale} (\omega^2-\nu^2)q_i=\nu^2{\cal K}_0\sum_{j\neq i}
\frac{1}{|\xi_i^{(0)}-\xi_j^{(0)}|^3}(q_i-q_j)
\end{equation} 
where we have introduced the dimensionless constant
\begin{equation}
\label{K:0} 
{\cal K}_0=\frac{2Q^2}{m\nu^2 L(N)^3}=\frac{2}{3N\log N}.
\end{equation} 
If the number of ions is large ($N\gg 1$), for the long-wavelength modes
one can approximate the chain by a continuous distribution of charges.
In this limit, $\xi$ is a 
continuous variable varying in the interval $(-1,1)$, while the displacement 
$q_i= q(\xi_i)$ is a continuous function, here denoted by $q(\xi)$. Then,
Eqs.~(\ref{Ax:Rescale}) take the form
\begin{eqnarray}
\label{Eq:Cont}
(\omega^2-\nu^2)q(\xi) =\frac{3}{4}\nu^2{\cal K}_0 N I[\xi,q(\xi)]
\end{eqnarray}
where 
\begin{eqnarray}
\label{I}
I[\xi,q(\xi)]
&=&\int_{-1}^{\xi-a(\xi)}
{\rm d}\xi'\frac{n(\xi')}
{(\xi-\xi')^3}(q(\xi)-q(\xi'))\\
& &+\int_{\xi+a(\xi)}^1 {\rm d}\xi'\frac{n(\xi')}{(\xi'-\xi)^3}
(q(\xi)-q(\xi'))
\nonumber
\end{eqnarray}
while $n(\xi)=1-\xi^2$ is the density of charges normalized to 4/3.
Equations~(\ref{Eq:Cont}) and~(\ref{I}) 
are valid away from the edges of the chain and for
long--wavelength excitations, where the continuum
approximation is reasonable. The continuum limit for Eqs.~(\ref{Eq:y}) 
gives
\begin{equation}
(\nu_t^2-\omega^2)y(\xi) =\frac{3}{8}
\nu^2{\cal K}_0 N I[\xi,y(\xi)].
\label{Eq:Cont:y}
\end{equation}
A similar type of equations is obtained for $z_i$. 
It is remarkable that the axial trap frequency
enters only as a prefactor on the right hand side of
Eqs.~(\ref{Eq:Cont}).
Consequently the axial eigenfrequencies are proportional to
$\nu$. The transverse eigenfrequencies, instead, do not show this behaviour,
as one can see from Eq.~(\ref{Eq:Cont:y}): Here, it is the quantity
$\omega^2-\nu_t^2$ which is proportional to $\nu^2$. 
The proportionality constants 
depend on the modes and will be calculated in what follows within some 
approximations. The results are independent of the charge $Q$ and of the mass
$m$. 

According to Eqs.~(\ref{Eq:Cont}-\ref{Eq:Cont:y}), the secular problem 
consists of solving the eigenvalue equation
\begin{equation}
\label{sec:I}
I[\xi,w^{(n)}(\xi)]=\lambda_{n}w^{(n)}(\xi)
\end{equation}
where $w^{(n)}(\xi)$ can be the axial or the transverse modes, that 
satisfy the orthogonality relation 
\begin{equation}
\label{Orthohonal}
\int_{-1}^1{\rm d}\xi n(\xi)
w^{(n)}(\xi)^{*}w^{(m)}(\xi)=\delta_{n,m}.
\end{equation}
By partial integration Eq.~(\ref{I}) can be written as the sum of two terms
$$I=I_0+\Delta I,$$
where $I_0$ contains the contributions of the ions around the point 
$\xi$, while the term $\Delta I$ is determined by the value of the density and
of the eigenmode function $q(\xi)$ and their derivatives at the end points
of the chain. In appendix A we derive the explicit form of the two terms and
discuss their order of magnitude. 
For the long wavelength modes and at the points $\xi$ sufficiently far away
from the chain end-points $\xi=\pm 1$, we find $I\approx I_0$ where 
\begin{eqnarray}
I_0[\xi,w(\xi)]
&=&\left(\log a(\xi)-\frac{3}{2}\right)
(n(\xi)w^{\prime\prime}(\xi)+2 n^{\prime}(\xi)w^{\prime}(\xi))\nonumber\\
& &+{\rm o}(a(\xi)),
\label{I0:Leading}
\end{eqnarray}
and $a(\xi)$ is an infinitesimal quantity of the order $1/N$.
Using~(\ref{Abstand}) and~(\ref{Gauss})
in~(\ref{I0:Leading}) we obtain
\begin{eqnarray}
I_0[\xi,w(\xi)]
=-\left(\log N +\log \frac{3}{4}+\log(1-\xi^2)+\frac{3}{2}\right)
J[w(\xi)]\nonumber\\
\label{I0:res}
\end{eqnarray}
where
\begin{equation}
\label{Jacobi}
J[w(\xi)]=(1-\xi^2)w^{\prime\prime}(\xi)-4\xi w^{\prime}(\xi)
\end{equation}
In the limits of validity of Eq.~(\ref{I0:Leading}) and for 
$N$ sufficiently large, such that $\log N\gg 1$, Eq.~(\ref{I0:res}) can be
approximated by the leading order in $\log N$
\begin{equation}
I_0[\xi,w(\xi)]\approx -\log N J[w(\xi)]
\end{equation}
and the eigenvalue equation~(\ref{sec:I}) reduces to
\begin{eqnarray}
\label{Jacobi:2}
J[w^{(n)}(\xi)]=\tilde{\lambda}_n w^{(n)}(\xi)
\end{eqnarray}
where $\lambda_{n}=-\log N\tilde{\lambda}_n$. Equation~(\ref{Jacobi:2}) 
is the differential equation fulfilled by the Jacobi polynomials
$P^{1,1}_{\ell}(\bar{x})$ at the eigenvalues~\cite{Abramowitz} 
\begin{equation}
\label{Jacobi:3}
\tilde{\lambda}_n=-\ell(\ell+3)
\end{equation}
with $\ell=0,1,\ldots$ and $n=\ell +1$.
After substitution of~(\ref{Jacobi}-\ref{Jacobi:3})
into~(\ref{sec:I}), the eigenfrequencies of the
axial excitations are found from~(\ref{K:0}-\ref{Eq:Cont}) and take the form
\begin{equation}
\label{omega:ax}
\omega_n^{\|~{\rm Jac}}=\nu\sqrt{\frac{n(n+1)}{2}}.
\end{equation}
Analogously, the eigenfrequencies of the transverse modes are obtained from 
Eq.~(\ref{Eq:Cont:y}) with~(\ref{sec:I}), resulting in
\begin{equation}
\label{omega:tr}
\omega_n^{\perp~{\rm Jac}}=\sqrt{\nu_t^2-\frac{(n-1)(n+2)}{4}\nu^2}.
\end{equation}
It should be noted that the solutions of (\ref{Jacobi})
for $\ell=0$ and $\ell=1$ are exact solutions of the original
problem~(\ref{Ax:Rescale}):  
The corresponding eigenmodes 
$w^{(1)}(x)=$constant (center of mass) and $w^{(2)}(x)\propto x$ (stretch
mode), which are the continuum limits of the eigenmodes we have found for the 
discrete case, are in fact Jacobi polynomials. The result for
the center of mass is obvious. The exact result for the stretch mode can be 
understood, noting that $P_1^{1,1}(\xi)$ has only one node, whose position  
coincides with the center of the chain $\xi=0$ and thus with the 
symmetry center for reflections. Hence, its position is
independent of the number of ions in the chain, and in particular it is
independent of whether the ions
distribution is discrete or continuous. 

It is remarkable that the dispersion
relation~(\ref{omega:ax}) coincides with a specific one dimensional 
limit of a three-dimensional continuum mean-field theory, like the one developed
in~\cite{Dubin91}, although there is no obvious justification for this. 
The two limiting cases, the uniform
spheroidal fluid of~\cite{Dubin91} and the case of $N$ strongly-coupled
oscillators investigated in this work, seem to provide the same 
axial eigenfrequencies 
in the long wavelength regime and in the limit $N\gg 1$. This
result is intriguing, especially if put in connection with theory of cold gases
in low dimensions, where different dispersion relations are obtained  
depending on the assumption on the type of mean-field
interaction~\cite{Menotti02,Stringari98}, and will be object of future
investigations. 

\begin{center}
\begin{figure}
\epsfxsize=0.3\textwidth
\epsffile{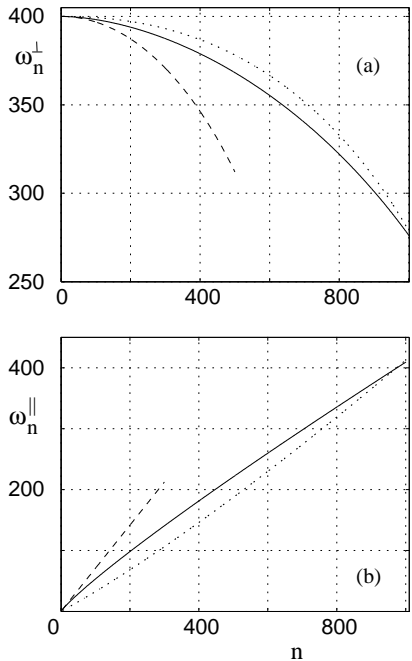}
\caption{(a) Transverse and (b) axial spectrum of eigenfrequencies
(in units of $\nu$) for a chain of $N=1000$ ions
and with $\nu_t=400\nu$. The solid line corresponds to the numerical
solution of Eqs.~(\ref{Eq:ax}-\ref{Eq:z}) with~(\ref{Eq:Discrete}). 
The dashed line shows 
(a) $\omega_n^{\perp~{\rm Jac}}$, (b) $\omega_n^{\|~{\rm Jac}}$: These curves
have been truncated, as they do not correctly reproduce the short wavelength
eigenmodes. The dash-dotted line gives the
spectra evaluated using the method of Dyson~\cite{Dyson53} implemented in subsection~\ref{Sec:Tr}.}  
\label{Fig:1}
\end{figure}
\end{center}
\begin{center}
\begin{figure}
\epsfxsize=0.5\textwidth
\epsffile{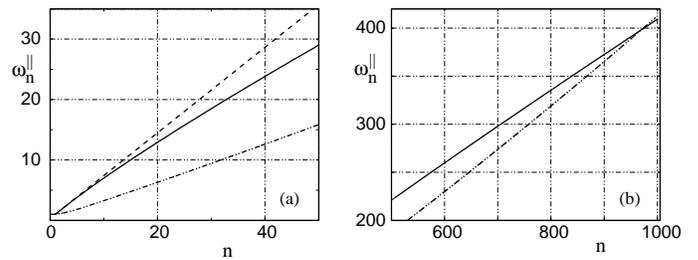}
\caption{
(a) Long--wavelength excitations and (b)
Short--wavelength excitations of the axial spectrum of eigenfrequencies. 
Same notation and parameters
as in Fig.~\ref{Fig:1}.} 
\label{Fig:2}
\end{figure}
\end{center}

Figure~\ref{Fig:1} presents the comparison between the spectrum of
eigenfrequencies, obtained by numerically diagonalizing the matrix
corresponding to Eqs.~(\ref{Eq:ax}-\ref{Eq:z}) for 1000
ions, and the formula~(\ref{omega:ax}) and (\ref{omega:tr}) derived 
for the axial and the transverse modes.
Figure~\ref{Fig:2}(a) exhibits the part of
the spectrum with the long--wavelength axial modes: Here, one sees that
Eq.~(\ref{omega:ax}) approximates well the lowest part of the axial spectrum,
where the limit of continuous charges distribution is reasonable. The
short--wavelength modes are not reproduced correctly by~(\ref{omega:ax}).
These are better evaluated by using a more
proper approximation for this regime, and will be discussed in the next
section.

We remark that, 
apart for the first two eigenmodes, the Jacobi polynomials describe the 
eigenmode excitation at leading order in $\log N$ and near the center of the
chain, where the  
interparticle separation is of order $1/N$ and the distribution of charges can
be treated as a continuum for sufficiently long wavelengths. The continuum 
approximation fails at the edges, where the interparticle spacing is
significantly larger
and Eq.~(\ref{Gauss}) is not meaningful. In particular, Eq.~(\ref{Length})
gives the upper bound for the chain length, which would be obtained in the
limit of $N\gg 1$ particles. Hence, a reasonable boundary condition is to
assume that the  
eigemodes and their derivatives vanish at $x=\pm L$, where there are no 
charges and hence the energy density is zero. The solution~(\ref{Jacobi}) 
taken at the center of the chain neglects the charges at the edges, on the 
basis of the observation that there the number of ions
is much smaller than at the center, and their contribution to the
integral~(\ref{I}) can therefore be neglected. 

\begin{center}
\begin{figure}
\epsfxsize=0.5\textwidth
\epsffile{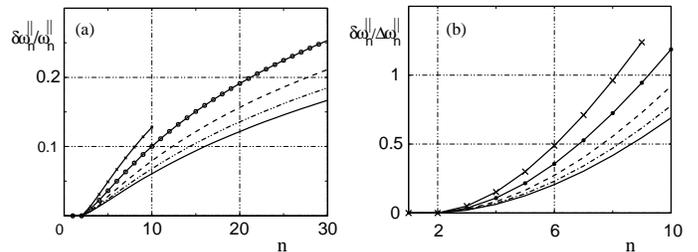}
\caption{(a) $\delta\omega_n^{\|}/\omega_n^{\|}$
and (b) $\delta\omega_n^{\|}/
  \Delta \omega_n^{\|}$
as a function of $n$, where
$\delta\omega_n=\omega_n^{\|{\rm Jac}}-\omega_n^{\|}$ and
$\Delta\omega_n^{\|}=\omega_{n+1}^{\|}-\omega_n^{\|}$.
The frequencies $\omega_n^{\|}$ are obtained by solving
numerically~(\ref{Eq:ax}) with~(\ref{Eq:Discrete}).
From top to bottom: $N=10, 50, 200, 500, 1000$.}
\label{Fig:0}
\end{figure}
\end{center}

The evaluation of the correction to the results (\ref{omega:ax})
and~(\ref{omega:tr}) should be done in perturbation theory in the 
parameter $1/\log N$, following an analogous procedure to the one applied
in~\cite{Dubin97} for evaluating 
the correction to the density of ions~(\ref{Gauss}) and to the equilibrium 
length~(\ref{Length}). In practice,
this expansion has a very slow convergence, and does 
not allow for a simple analytical expression. Nevertheless, the comparison 
with the spectrum evaluated numerically, by solving 
equations~(\ref{Eq:Discrete}) and (\ref{Eq:ax}-\ref{Eq:z}), shows that
Eq.~(\ref{omega:ax}) and~(\ref{omega:tr}) give already a good estimate of
the eigenfrequencies for a chain of 10 ions, as it can be seen in
Fig.~\ref{Fig:0}(a), where the relative deviation of the frequency given
by Eq.~(\ref{omega:ax}) from the numerical result is plotted for 
chains with different number of ions. This agreement is in general valid,
respectively, 
for the axial low modes and the transverse high modes, 
and exhibits a very slow improvement as the number of ions
in the chain increases, due to the slow convergence of the $1/\log N$ 
expansion. 

\subsection{The density of states in the short-wavelength limit}
\label{Sec:Tr}

Simple physical considerations show that the short-wavelength eigenmodes are
characterized by relatively large displacements of the ions around the center
of the chain, while the ions at the edges nearly do not move. In
fact, the interparticle distance is minimal in the middle of 
the chain, and for $N\gg 1$ it is of order $1/N$, while it is consistently
larger at the edges. Hence, a wave cannot propagate in a region where the
interparticle spacing is larger than the wavelength. 
Here, we can make some
simplifying assumptions in solving
the eigenvalue equations (\ref{Eq:ax}-\ref{Eq:z}) 
for the short wavelength modes. In fact, in the center of the chain we expect
that the relevant contributions to the force
on an ion originates from the neighbouring charges, as 
nearby groups of charges move in opposite directions, resulting in
forces that mutually cancel. By this hypothesis, in~(\ref{Eq:ax}-\ref{Eq:z})
we keep only the nearest--neighbour interaction, so that the equations to
solve are
\begin{eqnarray}
\label{Eq:sol:1}
&&\ddot{q_i}+\nu^2q_i=\Lambda[q_i,q_{i\pm  1}]\\
&&\ddot{y_i}+\nu_t^2y_i=-\frac{1}{2}\Lambda[y_i,y_{i\pm  1}]\\
&&\ddot{z_i}+\nu_t^2z_i=-\frac{1}{2}\Lambda[z_i,z_{i\pm  1}]
\label{Eq:sol:2}
\end{eqnarray}
where the tridiagonal matrix $\Lambda$ is defined by its action
on the vector $(w_1,\ldots,w_{i-1},w_i,w_{i+1},\ldots,w_N)$ through 
\begin{equation}
\Lambda[w_i,w_{i\pm 1}]=\Lambda_{i}(w_{i+1}-w_i)
+\Lambda_{i-1}(w_{i-1}-w_i)
\end{equation} 
and $\Lambda_{i}=K_{i,i+1}/m$ where $K_{i,j}$ is given in~(\ref{Kij}). 
The matrix $\Lambda$ is symmetric, and we denote its characteristic
frequencies by $-\tilde{\omega}^2$. Following the derivation of
Dyson~\cite{Dyson53},
that is summarized in appendix B, the density of states 
$D(\tilde{\omega}^2)$ for $N\gg 1$ is found from the characteristic function
$\Omega(\theta)$ according to 
\begin{equation}
\label{D1z}
D(1/z)=-z^2{\rm Re}\left[\frac{1}{{\rm i}\pi}\lim_{\epsilon\to
      0}\frac{d\Omega}{d\theta}(-z+{\rm i}\epsilon)\right]
\end{equation}
while $\Omega(\theta)$ is explicitly evaluated by using the properties of 
antisymmetric matrices and takes the form
\begin{equation} 
\label{Characteristic:0}
\Omega(\theta)= 
\lim_{N\to\infty}\frac{1}{N}\sum_{j=1}^{2N-1}\log[1+\zeta(\theta,j)]. 
\end{equation}
Here, $\zeta(\theta,j)$ is the infinite continued fraction,
\begin{equation} \label{Cont:Frac}
\zeta(\theta,j)=\frac{\theta\tilde{\Lambda}_j} 
{1+\frac{\theta\tilde{\Lambda}_{j+1}} 
{1+\frac{\theta\tilde{\Lambda}_{j+2}}{...}}}=
\frac{\theta\tilde{\Lambda}_j}{1+\zeta(\theta,j+1) }
\end{equation} 
and $\tilde{\Lambda}_{2i-1}=\tilde{\Lambda}_{2i}=\Lambda_i$.
For $N\gg 1$ and around the center of the chain 
we may assume $\Lambda(\xi)$ to be a slowly-varying function of
the position $\xi$, such that
$\Lambda_{i+1}=\Lambda_i+\delta\Lambda_i$ and
$\delta\Lambda_i/\Lambda_i\ll 1$. This allows to evaluate explicitly 
$\zeta(\theta,j)$ at first order in
$\delta\Lambda_i$. For this purpose, we define
$\tilde{\Lambda}_{j-1}=\tilde{\Lambda}_{j}=\Lambda$, and
$\tilde{\Lambda}_{j+1}=\tilde{\Lambda}_{j+2}=\Lambda+\delta\Lambda$,
where $\delta\Lambda$ is the first order variation.
Furthermore, we denote $\zeta(\theta,j)=\zeta$ and assume 
$\zeta(\theta,j+2)=\zeta+\delta\zeta$, where
$\delta\zeta$ is a first order variation. We substitute 
these quantities into~(\ref{Cont:Frac}) keeping only the terms up to 
first-order and look for a consistent 
solution. The resulting equation is
\begin{equation} \delta\zeta(2\zeta+1)=\theta\delta\Lambda \end{equation}
which is integrated to 
\begin{equation} \label{Eq:3}
\zeta^2+\zeta-\theta\Lambda=0.
\end{equation} 
Here, we have taken
the integration constant to be zero since at the boundaries of the
chain $\Lambda\to 0$. The resulting solution has the form
\begin{equation} 
\zeta(\theta,\xi)=\frac{1}{2}\left[\sqrt{1+4\theta\Lambda(L\xi)}-1\right] 
\end{equation}
leading to the characteristic function
\begin{eqnarray}
\nonumber
&&\Omega(\theta)
=\lim_{N\to\infty}\frac{2}{N}\sum_{j=1}^{N-1}\log\left[\frac{1}{2}
\left(\sqrt{1+4\theta\Lambda(x_j)}+1\right)\right]\\
&&=3\int_{0}^{1}{\rm d}\xi~n(\xi)\log\left[\frac{1}{2}
\left(\sqrt{1+4\theta\Lambda(L\xi)}+1\right)\right]
\label{Characteristic:1}
\end{eqnarray}
where we have used the rescaled variable $\xi$ and the fact 
that the integrand is even 
in the interval $(-1,1)$. Equation~(\ref{Characteristic:1}) corresponds to
the continuum
limit of the discrete summation in Eq.~(\ref{Characteristic:0}), and it is
valid away from the edges for $N\gg 1$. Substituting~(\ref{Characteristic:1})
into~(\ref{D1z}) we obtain the equation for the density of states as a
function of the physical parameters,
\begin{eqnarray}
D(1/z)=
\frac{6z}{\pi}\int_{0}^{f(z)}{\rm d}\xi~n(\xi)\left[\frac{1}
{\sqrt{-1+4z\Lambda(L\xi)}}\right]
\label{Density:Modes}
\end{eqnarray}
where $f(z)=\sqrt{1-(1/4\Lambda_0 z)^{1/3}}$, while
\begin{equation}
\label{lambda:x}
\Lambda(L\xi)=\Lambda_0(1-\xi^2)^3,
\end{equation}
with 
\begin{equation}
\label{lambda:0}
\Lambda_0=\frac{{\cal K}_0}{a(\xi=0)^3}\nu^2\approx\frac{9}{32}\frac{N^2}{\log
  N}\nu^2
\end{equation}
at leading order in $\log N$.\\

\begin{center}
\begin{figure}
\epsfxsize=0.3\textwidth
\epsffile{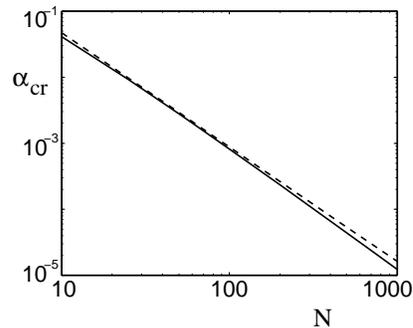}
\caption{$\alpha_{\rm cr}$ as a function of the number of ions. The
  solid curve gives the result~(\ref{alpha}). The
  dashed curve is a fit according to the function $cN^{-1.73}$, with $c=2.53$,
as calculated numerically
in~\protect\cite{Schiffer93} (see also~\protect\cite{LosAlamos00}).}
\label{Fig:Stable}
\end{figure}
\end{center}

Equation~(\ref{Density:Modes}) with $\tilde{\omega}^2=1/z$ gives the 
density of states of the short 
wavelength modes. The corresponding spectrum is shown in Fig.~\ref{Fig:1}.
The deviations from the numerical result are due to the
assumption of nearest-neighbour coupling, and are small in the
short-wavelength part of the spectrum, showing that Eq.~(\ref{Density:Modes})
provides a good approximation in this regime. In particular, this result
allows to evaluate explicitly the value of the maximal axial frequency 
$\omega_{\rm max}^{\|}$, the minimal transverse frequency
$\omega_{\rm min}^{\perp}$, and the spectrum of the eigenfrequencies
in their neighbourhood. The maximal axial frequency and the minimal transverse
frequency are found from the maximal value of $z$ for 
which the integrand in~(\ref{Density:Modes}) is real, corresponding to
$f(z)=0$. For larger values of $z$ the density of states vanishes. The
corresponding eigenvalue of the matrix $\Lambda$ is
$\tilde{\omega}^2=4\Lambda_0$. From~(\ref{Eq:sol:1}-\ref{Eq:sol:2}) one finds
$\omega_{\rm max}^{\|}=\sqrt{\nu^2+4\Lambda_0}$ and 
$\omega_{\rm min}^{\perp}=\sqrt{\nu_t^2-2\Lambda_0}$.
Therefore, the largest value of the axial 
frequency is determined by the largest
value of the spring constant, that is the value of the spring constant at
the center of the chain, and at leading order in $\log N$ is
\begin{equation}
\label{w:max}
\omega_{\rm max}^{\|}\approx\nu
\sqrt{1+\frac{9}{8}\frac{N^2}{\log N}}.
\end{equation}
Analogously, the smallest value of the transverse modes frequency is
\begin{equation}
\label{w:min:t}
\omega_{\rm min}^{\perp}\approx\sqrt{\nu_t^2-\frac{9}{16}\nu^2
\frac{N^2}{\log N}}.
\end{equation}
From Eq.~(\ref{w:min:t}) one sees that $\omega_{\rm min}^{\perp}$ can
vanish for certain values of $\nu,\nu_t$, and $N$.
We denote by
\begin{equation}
\label{nu:t:cr}
\nu_{t,{\rm cr}}=\sqrt{\frac{9}{16}\nu^2 \frac{N^2}{\log N}}
\end{equation} 
the value of the transverse frequency, such that for $\nu_t<\nu_{t,{\rm
    cr}}$ the linear chain is unstable with respect to excitations of the
    transverse vibrations. Using the notation introduced in~\cite{Schiffer93}
    we define the critical value of the aspect ratio between the trap
    frequencies $\alpha_{\rm cr}=\nu^2/\nu_{t,{\rm cr}}^2$. It takes the value 
\begin{equation}
\label{alpha}
\alpha_{\rm cr}=\frac{16}{9}\frac{\log N}{N^2}
\end{equation}
which fixes the condition on $\nu_t$ for the chain stability 
according to the inequality
$\nu_t^2>\alpha_{\rm cr}^{-1}\nu^2$, and it is in agreement with the analytical
estimate in~\cite{Dubin93}, which was obtained under different requirements. 
In Fig.~\ref{Fig:Stable} we
compare the result~(\ref{alpha}) with the relation $cN^{\beta_1}$,
with $\beta_1=-1.73$ and $c=2.53$, obtained in~\cite{Schiffer93} 
by fitting points calculated with molecular dynamics
simulations, and verified experimentally in~\cite{LosAlamos00}. Our result
reproduces approximately this curve. At $\alpha_{\rm cr}$ the
crystal undergoes a structural transition from a linear string to a zig-zag
configuration, as it has been observed in~\cite{MPQ}.
It is interesting to ask whether this structural transition can
be considered as a phase transition, as discussed 
in~\cite{Schiffer93,Dubin93,LosAlamos00}.
A systematic investigation in this direction 
requires a proper definition of the thermodynamic limit for this kind of
system. A natural thermodynamic limit, that will be discussed in the next
section, is one where as $N\to\infty$, the axial frequency $\nu\to 0$ so that
the density in the center is fixed. From~(\ref{Gauss}) and~(\ref{Length}) this
requires that the ratio $\nu^2N^2/\log N$ is kept constant. 
From~(\ref{nu:t:cr}) it
is clear that in this limit the critical transverse frequency tends to a well
defined value. The exploration of the properties of this transition in the
thermodynamic limit will be object of further
studies. Particularly interesting is the comparison with standard phase
transitions~\cite{Lubensky}. 

\subsection{The phonon-like approximation}

It is natural to introduce a phonon-like approximation for the eigenmodes of
Eqs.~(\ref{Eq:ax}-\ref{Eq:z}). In this approximation, a phonon-like solution
is superimposed by a slowly varying amplitude, that takes into account the
slow variation of the coupling strength $K_{i,j}$ of Eq.~(\ref{Kij}) as a
function of both $i$ and $j$. This approximation, that is outlined in appendix
C, is reasonable for a relatively wide range of long wavelength excitations
compared to the Jacobi polynomials solution, discussed in Sec.~\ref{Sec:Jac} (see
Fig.~\ref{Fig:Last}). It is inferior compared to the Jacobi polynomials in the
very long wavelenght limit, and it is a bad approximation for the short
wavelength regime. Therefore, the phonon-like approximation, that is natural
in condensed matter physics, is not an asymptotic approximation in the
thermodynamic limit $N\to\infty$, neither for the long wavelength nor for the
short wavelength part of the spectrum.

\section{Statistical Mechanics}

In this section we use the density of states, that was evaluated in the
preceding section, in order to derive the 
thermodynamic quantities of a linear crystal of $N$ ions. The linear chain
is assumed to be in the regime of stability and 
to be in equilibrium with a thermal bath at temperature $T$. The oscillations
around the classical equilibrium points are quantized using standard
procedures~\cite{James}. 
It should be noted that in this limit the quantum statistics of the atoms
are irrelevant: In fact, the single particle wavepacket is much smaller than
the interparticle distance~\cite{Javanainen}.
The dynamics of the crystal is thus
described by $3N$ harmonic oscillators of frequencies $\omega_n^{\|}$,
$\omega_n^{\perp}$, where the frequencies 
$\omega_n^{\perp}$ are doubly
degenerate. It is modeled by the Hamiltonian $\hat{H}$, obtained
from~$H$ after quantizing the eigenmodes of $H_{\rm har}$ in~(\ref{H:1}). Here,
$\hat{H}=H_0+\hat{H}_n$, where $H_0$ is the ground state energy, 
$$H_0=V_0+\frac{\hbar}{2}\sum_{n=1}^N\left(\omega_n^{\|}+2\omega_n^{\perp}
\right)$$
while $\hat{H}_n$ describes the contribution of the collective excitations
$$\hat{H}_n=\sum_{n=1}^N\hbar\omega_n^{\|}\hat{N}_n^{\|} +
\sum_{n=1}^{N}\hbar\omega_n^{\perp}(\hat{N}_{n,y}^{\perp}+\hat{N}_{n,z}^{\perp})$$ 
where $\hat{N}_n^{\|}$, $\hat{N}_{n,y}^{\perp}$, $\hat{N}_{n,z}^{\perp}$ 
are the operators counting the
number of excitations. The term $V_0$ corresponds to the classical 
minimum energy of the Coulomb crystal. For an infinite chain, it is obtained
by minimizing the classical energy with respect to the length $L$ using the
density of charges in Eq.~(\ref{Gauss}), and it is evaluated to
be~\cite{Dubin97}
\begin{equation}
\label{Energy:Zero}
V_0=\frac{3}{10}m\nu^2NL(N)^2
\end{equation}
where $L(N)$ is given in~(\ref{Length}).

We remark that we investigate the thermodynamic quantities for crystals
characterized by a finite number of particles $N$ and finite axial frequencies 
$\nu$, i.e.\ crystals of finite size, that may be close to experimental
situations. It is however instructive to consider the definition of the
thermodynamic limit for this kind of system characterized by strong
correlations, where the effect of the 
charges at the edges cannot be neglected a priori in evaluating
the statistical properties.
Here, the thermodynamic limit can be appropriately 
defined by assuming constant interparticle spacing (thus constant 
linear density) at the center of the chain $x=0$, namely requiring
$a_L(0)$ to be constant. Denoting by $a_0=a_L(0)$, 
it scales as $$a_0=g\left(\frac{\sqrt{\log N}}{\nu N}\right)^{2/3},$$ 
where $g=(3Q^2/m)^{1/3}$. This
requirement corresponds to a vanishing axial frequency, according to $\nu\sim
\sqrt{\log N}/N$, as $N\to\infty$. With this definition,
in the thermodynamic limit 
the maximum axial frequency~(\ref{w:max}) and the critical transverse
frequency~(\ref{nu:t:cr}) are independent of $N$ and of $\nu$, taking the
values $\omega_{\rm max}^{\|}=3 (g/2a_0)^{3/2}$ and $\nu_{t,{\rm cr}}=3/4 
(g/a_0)^{3/2}$.
In the following, we derive the thermodynamic quantities for ion chains of
finite size characterized by a fixed and finite value of the 
number of particles $N$ and of the 
axial frequency $\nu$, and discuss how these quantities behave when
we take the specific thermodynamic limit that was defined above. 

Assuming thermal equilibrium with the bath, 
the state of the system at constant number of ions $N$
is described by the density matrix of the canonical ensemble 
\begin{equation}
\rho=\frac{1}{Z}\exp\left(-\beta H\right) 
\end{equation} 
where $\beta=1/k_BT$ and $Z$ is the partition function, 
$$Z=\exp\left(-\beta H_0\right)\prod_n
[1-\exp(-\beta \hbar\omega_n)]^{-1}$$ 
which determines the Helmoltz free energy $$F=-k_BT\log Z.$$

\begin{center}
\begin{figure}
\epsfxsize=0.3\textwidth
\epsffile{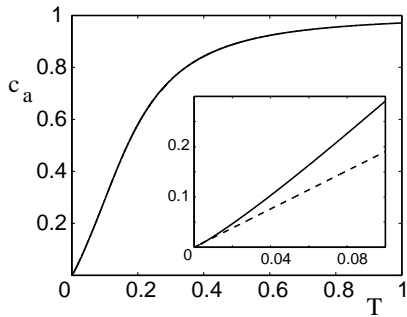}
\caption{Specific heat $c_a=C_a/N$, in units of $k_B$, 
as a function of the temperature $T$, in units of $\hbar\omega_{\rm
  max}^{\|}/k_B$. Here, $N=1000$ and $\omega_{\rm
  max}^{\|}\ll \omega_{\rm min}^{\perp}$. This would correspond to an
  experiment with $\nu=1\pi\times 1$ kHz and
$\nu_t=2\pi\times 4$ MHz. The inset
shows the low-temperature behaviour, the dashed line is the estimated slope,
with $\tilde{c}=\sqrt{8}\zeta(2)k_B^2/\hbar\nu$ (see text).}
\label{Fig:cV}
\end{figure}
\end{center}

We identify the thermodynamic variables with $T$, the temperature, $N$, the
number of atoms, and $\nu$, 
the axial trap frequency, whose variation corresponds to a variation in the 
length of the chain~\cite{DubinRMP}. These are not
a complete set of thermodynamic variables, but they fully determine the state
of the crystal for the thermodynamic quantities we investigate in the
following. In particular, we take $\nu_t$ as constant and
assume $\omega_{\rm max}^{\|}\ll \omega_{\rm
  min}^{\perp}$, i.e.\ that there is a large gap between axial and
transverse excitations. In this limit, we 
consider temperatures $T$ such that $k_B T\ll \hbar\omega_{\rm min}^{\perp}$. 
In this regime the transverse modes can be considered frozen, hence the 
contribution due to their excitations to the crystal's thermodynamic 
properties can be neglected, and the dynamics of the system is
one-dimensional. The thermal energy of the crystal is given by
\begin{equation}
U_{\rm th}=\langle \hat{H}_n\rangle=\sum_n \frac{\hbar\omega_n^{\|}}
{\exp(\beta \hbar\omega_n^{\|})-1}.
\end{equation}
The heat capacity $C_a=\partial U_{\rm th}/\partial T|_{\nu,N}$ is
\begin{eqnarray} 
\label{cL:0}
C_a=\frac{\partial}{\partial T}\sum_n\frac{\hbar\omega_n^{\|}}
  {\exp(\beta \hbar\omega_n^{\|})-1}.
\end{eqnarray}
The behaviour at high
temperatures, such that $k_B T\gg\hbar\omega_{\rm max}^{\|}$ (but $k_B T\ll 
\hbar\omega_{\rm min}^{\perp}$), is given by the Dulong--Petit 
law $C_a=Nk_B$ as is clear from~(\ref{cL:0}). 
On the other hand, for $k_BT\ll \hbar\nu$ all modes are
frozen and the energy of the chain is the zero point energy $H_0$.
For large number of particles $N\gg 1$ we can approximate the sum in~(\ref{cL:0}) by the integral 
\begin{eqnarray*}
C_a\sim \frac{\partial}{\partial T}
\int_{\nu}^{\omega_{\rm max}^{\|}} {\rm d}\omega
g(\omega)\frac{\hbar\omega}{\exp(\beta \hbar\omega)-1}
\end{eqnarray*}
where $g(\omega)=\partial n/\partial \omega$ is the density of states. In
particular, at temperatures such that the contribution of the 
long-wavelength excitations to the sum is predominant and is given
by~(\ref{omega:ax}), the resulting density of states is
$$g(\omega)\Bigl|_{\rm
  low~T}=\frac{1}{\nu}\frac{4\omega}{\sqrt{8\omega^2+\nu^2}}$$ 
and the heat capacity is given by the integral
\begin{equation}
C_a\bigl|_{\rm low~T}\sim \frac{\sqrt{2}k_B^2}{\hbar\nu}
\frac{\partial}{\partial T}T^2
\int_{x_0}^{\infty} {\rm d}x \frac{1}{{\rm e}^{x}-1}\frac{x^2}
{\sqrt{x^2+x_0^2/8}}
\label{cV:1}
\end{equation}
where we have defined $x_0=\beta\hbar\nu$. Hence, 
the integrand and the integration limits depend on the temperature through
$x_0$. In this regime and
for $k_BT\gg \hbar\nu$ we can set $x_0\sim 0$ in (\ref{cV:1}),
and recover the result $C_a=\tilde{c} T$, with
$\tilde{c}=\sqrt{8}\zeta(2)k_B^2/\hbar\nu$, where $\zeta(2)$ is the Riemann's
zeta function. Therefore, for the considered regime the heat capacity is
proportional to the temperature, which is a characteristic behaviour 
encountered in a one-dimensional Debye crystal~\cite{Ashcroft}. 
In Fig.~\ref{Fig:cV} the specific heat $c_a=C_a/N$ for $N=1000$ ions
is plotted as a function of the temperature $T$. In the inset, the low
temperatures behaviour is shown, and compared with the curve
$\tilde{c}T/N$ estimated above. The figure shows that the evaluated
behaviour, valid in the asymptotic limit of an infinite number of ions,
provides a reasonable description of the specific heat at low temperatures.

It is remarkable that the heat capacity in~(\ref{cV:1})
scales like $C_a\sim 1/\nu$. The specific heat per particle $c_a=C_a/N$
behaves thus like $c_a\sim 1/N\nu$ at low temperatures, 
and in the thermodynamic limit it vanishes as $c_a\sim 1/\sqrt{\log N}$. 
It thus depends on the number of ions, and this is a manifestation of the 
deviation from extensivity of the system's behaviour. 
Note that the relative energy
fluctuations are of the order $(\sqrt{\log N}/N)^{1/2}$,
therefore the usual equivalence of
ensembles holds. 

The pressure $P$ in the axial direction is defined as 
\begin{eqnarray}
P=-\frac{\partial F}{\partial L}\Bigl|_{T,N}
\label{pressure}
\end{eqnarray}
and it is the variation of the free energy with the length of the string 
at constant $N$ and $T$. Under these conditions, the length of the crystal is
varied by changing the axial trap frequency $\nu$, according to
\begin{equation}
\frac{\partial L}{\partial \nu}\Bigl|_{T,N}=-\frac{2L}{3\nu}
\end{equation}
as obtained from~(\ref{Length}). 
Substituting the explicit form of the free energy into~(\ref{pressure}), 
we find
$$P=P_0+P_T,$$
where 
\begin{eqnarray}
P_0
=-\frac{\partial H_0}{\partial L}\Bigl|_{T,N}=\frac{V_0}{L}+\frac{3\hbar}{4L}
\sum_n\left[\omega_n^{\|}+2\omega_n^{\perp}
\left(1-\frac{\nu_t^2}{\omega_n^{\perp 2}}\right)\right]\nonumber\\
\label{pressure:0} 
\end{eqnarray}
is the pressure at zero-temperature, and 
$P_T$ gives the contribution of the excitations, 
\begin{eqnarray}
P_T=\frac{3}{2L}U_{\rm th}.\label{pressure:T}
\end{eqnarray}
In deriving (\ref{pressure:0}) and (\ref{pressure:T}) 
we have used $\partial V_0/\partial L=-V_0/L$ that results
from~(\ref{Energy:Zero}), and the relations 
\begin{eqnarray*}
&&\frac{\partial}{\partial \nu}\omega_n^{\|}=\frac{\omega_n^{\|}}{\nu}\\
&&\frac{\partial}{\partial
  \nu}\omega_n^{\perp}=\frac{\omega_n^{\perp}}{\nu}\left[1 - \left(
    \frac{\nu_t}{\omega_n^{\perp}} \right)^2 \right] 
\end{eqnarray*}
that are obtained from the functional dependence of
$\omega_n^{\|}$ and $\omega_n^{\perp}$ on $\nu$, as can be extracted
from~(\ref{Eq:Cont}) and~(\ref{Eq:Cont:y}).
Note that a variation of the axial trap frequency implies also a variation of
the transverse eigenfrequencies, which give a contribution of opposite sign 
to the pressure, as it is obvious from~(\ref{pressure:0}). However, the
contribution due to the quantum mechanical zero--point energy is
very small compared to the classical term $V_0/L$. 
Therefore, $P_0$ is dominated by $V_0/L$ and, using
(\ref{Energy:Zero}), in the thermodynamic limit it scales as $P_0\sim\log
N$. The term $P_T$ depends on the temperature. 
For low temperatures, such that $k_BT\gg\hbar\nu$, it scales as
$P_T\sim 1/L\nu\sim 1/\sqrt{\log N}$. At high temperatures, in the
Dulong-Petit regime, $P_T$ does neither depend
on $N$ nor on $\nu$. In the regime where the chain is thermally stable,
which we consider here,
$V_0\gg U_{\rm th}$ giving $P\approx P_0$. Thus, the pressure is dominated by
the zero-temperature contribution and in the thermodynamic
limit $P\approx P_0\sim\log N$. 
A useful relation for the following discussion is 
\begin{equation}
\frac{\partial P}{\partial T}\Bigl|_{L,N}=\frac{3}{2L}C_a
\label{for:alpha}
\end{equation}

The isothermal compressibility $\kappa_T$ is evaluated from the pressure 
according to
\begin{eqnarray}
\label{kappa:T}
\frac{1}{\kappa_T}=B
=-L\frac{\partial P}{\partial L}\Bigl|_{T,N}
\end{eqnarray}
where $B$ is the bulk modulus. Using (\ref{pressure:0}), 
(\ref{pressure:T}) in (\ref{kappa:T}) we find
\begin{eqnarray}
\kappa_T
=\left[-L\frac{\partial P_0}{\partial L} +\frac{3}{4L}(
5U_{\rm th}-3C_aT)\right]^{-1}
\end{eqnarray}
In the thermodynamic limit the bulk modulus $B$ is
dominated by the zero-temperature contribution $-L\partial P_0/\partial L$,
that in turn is dominated by the term $-L\partial V_0/\partial L\sim V_0/L\sim
\log N$. Therefore, 
$B\sim \log N$ and the compressibility $\kappa_T$ vanishes as $1/\log N$. 

\begin{center}
\begin{figure}
\epsfxsize=0.3\textwidth
\epsffile{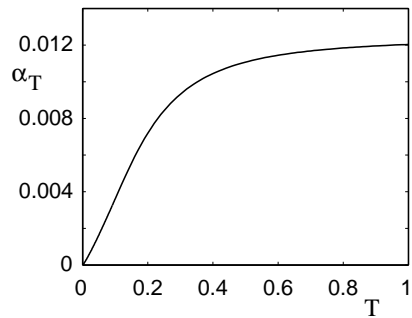}
\caption{Coefficient of
  thermal expansion $\alpha_T$, in arbitrary units, as a function of 
$T$, in units of
$\hbar\omega_{\rm max}^{\|}/k_B$. 
Here, $N=1000$ and $\omega_{\rm max}^{\|}\ll \omega_{\rm min}^{\perp}$.
This would correspond to an experiment with $\nu=1\pi\times 1$~kHz and
$\nu_t=2\pi\times 4$~MHz. For these
parameters and Berillium atoms, $T$ is in units of the Debye temperature
$\Theta_D$, with $\Theta_D=\hbar\omega_{\rm max}^{\|}/k_B
\approx 20\mu$K, and $\alpha_T$ in units of $10^{-5}\mu$K$^{-1}$.}
\label{alphaT}
\end{figure}
\end{center}

The coefficient of thermal expansion $\alpha_T$ can be evaluated from
the knowledge of $\kappa_T$ and $C_a$ according to~\cite{Ashcroft}
\begin{eqnarray}
\label{alpha:T:0}
\alpha_T=\frac{1}{L}\frac{\partial L}{\partial T}\Bigl|_{P,N}
=-\frac{1}{L}\frac{\partial P/\partial T|_L}{\partial P/\partial L|_T}
=\frac{3}{2L}\kappa_T C_a
\end{eqnarray}
where we have used~(\ref{for:alpha}). Since the
compressibility is dominated by the zero temperature term, the behaviour of
the coefficient of thermal expansion $\alpha_T$ as a function of $T$ 
is determined by the heat capacity: Linear dependence at
low temperatures, and saturation at high temperatures.
At low temperatures $C_a\sim 1/\nu$, and
as the thermodynamic limit is approached
$\alpha_T\sim (\log N)^{-3/2}$. At higher 
temperatures, when the heat capacity manifests the Dulong--Petit behaviour
$C_a=Nk_B$, the coefficient of thermal expansion vanishes like 
$\alpha_T\sim 1/\log N$. In Fig.~\ref{alphaT} the coefficient of
thermal expansion for a chain of 1000 ions is presented. 
Calculations made with different numbers of ions,
taking the trap frequency $\nu$ such that the linear density at the center of
the chain remains constant, show that $\alpha_T$ decreases with $N$. 
The numerical results are consistent with the behaviour we expect
in different temperature regimes, according to the above considerations. 
In particular, for finite $N$ it is significantly different from zero.
This is in contrast to the behaviour found for uniform harmonic solids, where
the coefficient of thermal expansion vanishes~\cite{Ashcroft}. 

The thermodynamic quantities of a linear string of charges
confined by a harmonic trap are affected by the way the thermodynamic
limit is taken. Nevertheless, the behaviour of the system is intrinsically
non-extensive. The non-extensitivity is due to the strong correlation and the
dimensionality of the crystal, which determine a regime where the correlation
energy, associated with the discreteness of the individual charges, cannot be
neglected~\cite{Dubin97}. It manifests in particular in the $\log N$
terms appearing in the thermodynamic quantities. 
A representative example is the dependence of the specific heat per
particle on $N$. 

We finally remark on the thermal stability of the chain. The derivation
presented in this section, in fact, relies on the assumption that the thermal
excitations do not affect the stability of the system and thus the validity of
the physical model we are considering, namely of ions oscillating around their
equilibrium positions. This condition is equivalent to the
statement that the thermal energy at the considered temperatures is
considerably 
lower than the equilibrium energy, and the displacements are much smaller
than the respective interparticle distance. Hence, 
this condition is valid when 
$$\frac{Q^2}{a_0}\log N\gg k_BT.$$
This relation is amply satisfied for
the parameters of the numerical evaluation of the
thermodynamic quantities considered in this section.  

\section{Summary and Conclusions}
We have investigated the dynamics of a linear crystal excitations with 
Coulomb interaction and in presence of an external potential.
We have derived an analytical formula for the density of states
at long- and short-wavelengths. In the long-wavelength part of the spectrum,
we have calculated analytically the eigenmodes and
eigenfrequencies. The eigenmodes and eigenfrequencies for
the center-of-mass and the first excitation of the axial and of the
transverse motion are exact and independent of the number of ions. Apart for these modes, 
the results we derive are valid in the limit of infinite number of
ions. Nevertheless, they already give a good description of the spectrum of
excitations of chains of dozens of ions. Using our results we study the
statistical mechanics and thermodynamics of the linear chain. 

Our derivation allows
to find an analytical formula for the critical transverse frequency required
for determining the stability of the linear chain. It agrees with the
analytical estimate by~\cite{Dubin93}, that was obtained under different
assumptions, and it is consistent with the
formula fitted from numerical data~\cite{Schiffer93} and verified
experimentally~\cite{LosAlamos00}. It was suggested that this instability of
the ion chain, resulting in a transition 
from a linear to a zig-zag equilibrium configuration, can
be treated as a phase transition~\cite{Schiffer93,LosAlamos00}. In future
works we will explore, using the formulation developed in this work, whether
the thermodynamic quantities exhibit singularities characteristic of phase
transitions~\cite{Lubensky}. This system differs from systems that are
traditionally studied in the framework of statistical physics, since it is not
extensive. In particular the specific heat per particle depends on the number
of ions. 

The results presented in this work show a statistical mechanics approach
applied to a strongly-correlated mesoscopic system.
They contribute to
the on-going research on low-dimensional cold gases~\cite{Stringari98,BEC1d} 
and may be relevant to studies of the quantum dynamics of many--particle 
Coulomb systems like ion crystals in storage rings~\cite{Habs}
and cold neutral plasmas~\cite{GallagherReview,Rolston}, 
which at sufficiently low
temperatures are predicted to crystallize~\cite{Pattard}. 
The connection with
Wigner crystals, where the quantum statistics may play a relevant
role~\cite{Schulz}, will be explored. 

Moreover, the spectra of excitations here evaluated
are relevant for the implementations of quantum logic with
ion traps, 
where the knowledge of the long-wavelength modes is important for the
realization of logic gates~\cite{Cirac95,Steane,QLogic:2,QLogic:3}, as well as
realization of solid-state models~\cite{Porras}.

\acknowledgements
It is our great pleasure to thank
Andreas Buchleitner for hospitality at the 
Max-Planck-Institut f\"ur Komplexe Systeme in
Dresden, where most research was done. S.F.\ is grateful to
Wolfgang Schleich and the department of quantum physics in Ulm for
hospitality. G.M.\ acknowledges the kind hospitality of the Institute for
Theoretical Physics at Technion in
Haifa. The authors
acknowledge stimulating discussions and helpful comments
of T.\ Antonsen, M.\ Drewsen, J.\ Eschner, D.\
Habs, P.\ Kienle, E.\ Ott, R.\ Prange, T.\ Sch\"atz, J.\ Schiffer, 
E.\ Shimshoni, and H.\ Walther. We thank the referee for bringing to our
attention the article in~\cite{Dubin&Schiffer}. 

This work was partly supported
by the US-Israel Binational Science Foundation (BSF), by the
Minerva Center of Nonlinear Physics of Complex Systems,
by the Institute of Theoretical Physics 
at the Technion, and by the EU-networks QUEST and QGATES.

\begin{appendix}

\section{Secular equation of the long--wavelength eigenmodes}

In this appendix we outline the derivation of the differential equation
(\ref{I0:res}-\ref{Jacobi}) from Eq.~(\ref{I}), and estimate the correction 
to the solution we find. By integration by parts, Eq.~(\ref{I}) takes the form
$$I=I_0+I_1+I_2$$
where
\begin{eqnarray}
\label{I_0}
I_0&=&\frac{1}{2a(\xi)^2}\left(f_{\xi}(\xi-a)+f_{\xi}(\xi+a)\right)\\
   & &-\frac{1}{2a(\xi)}\left(f_{\xi}^{\prime}(\xi-a)
-f_{\xi}^{\prime}(\xi+a)\right) \nonumber\\
   & &-\frac{1}{2}\log a(\xi)\left(f_{\xi}^{(2)}(\xi-a)+
f_{\xi}^{(2)}(\xi+a)\right)\nonumber\\
\label{I_1}
I_1&=&-\frac{1}{2}\left(\frac{f_{\xi}(-1)}{(1+\xi)^2}
+\frac{f_{\xi}(1)}{(1-\xi)^2}\right)\\
   & &+\frac{1}{2}\left(\frac{f{\xi}^{\prime}(-1)}{1+\xi}
-\frac{f_{\xi}^{\prime}(1)}{1-\xi}\right)\nonumber\\
   & &+\frac{1}{2}\left(f_{\xi}^{(2)}(-1)\log(1+\xi)
+f_{\xi}^{(2)}(1)\log(1-\xi)\right)\nonumber\\
\label{I_2}
I_2&=&\frac{1}{2}\Bigl[\int_{-1}^{\xi-a(\xi)}{\rm d}\xi'\log(\xi-\xi')
f_{\xi}^{(3)}(\xi')\\
   & &-\int_{\xi+a(\xi)}^1{\rm d}\xi'\log(\xi'-\xi)
f_{\xi}^{(3)}(\xi')\Bigr]\nonumber
\end{eqnarray}
and $f_{\xi}(\xi')=n(\xi')(w(\xi)-w(\xi'))$. Clearly the integral $I_0$
results of
the contribution of the ions around the ion at $\xi$. For long--wavelength modes
and away from the end-points $\xi=\pm 1$, 
the interparticle spacing $a(\xi)$ scales as $1/N$
and it can be treated as an infinitesimal quantity. 
Expanding~(\ref{I_0}) in $a(\xi)$
and keeping the leading order, we obtain
Eq.~(\ref{I0:Leading}). The integral~(\ref{I_1}) contains the
contribution of the end points of the chain.
Also, the integral~(\ref{I_2}) is dominated by the end points, as can be seen by
integrating repeatedly by parts, taking the indefinite integral of $\log x$,
and making the reasonable assumption that $f(x)$ is an analytic function on
the interval $(-1,1)$. 

By imposing the requirement that the density of ions and its 
derivatives vanish at the end-points, while $w(\xi)$ and its derivatives 
are bounded, the term $\Delta I=I_1+I_2$ is
much smaller than $I_0$. In particular, the
contributions at the end--points vanish, whereas the contributions to $I_2$
from the ions around the point $\xi$ are of order $a(\xi)\sim 1/N$.

\section{Characteristic function for the short--wavelength
  spectrum}

In this appendix we discuss the evaluation of the characteristic function 
for the eigenvalue problem in Eq.~(\ref{Eq:sol:1}) following the treatment
developed by Dyson~\cite{Dyson53}. 
The procedure can be
  immediately generalized to the eigenvalues of the transverse motion. 
Substituting $q_j(t)=\int {\rm d}\omega {\rm e}^{{\rm i}\omega t}
\tilde{q}_j(\omega)/2\pi$ 
into (\ref{Eq:sol:1}) and replacing $\tilde{q}_i(\omega)$
with $q_i$, we obtain
\begin{equation} 
-(\omega^2-\nu^2) q_j=\Lambda_{j}(q_{j+1}-q_j) +\Lambda_{j-1}(q_{j-1}-q_j) 
\end{equation}
The roots $-\tilde{\omega}^2$ of the matrix $\Lambda$ are
related to the eigenfrequencies $\omega$ by 
$$\omega=\sqrt{\tilde{\omega}^2+\nu^2}$$ 
These roots can be found by solving the second-order differential equation
\begin{equation} 
\label{Eq:2a} 
\ddot{\bar{q}}_j=\Lambda_{j}(\bar{q}_{j+1}-\bar{q}_j) 
+\Lambda_{j-1}(\bar{q}_{j-1}-\bar{q}_j) 
\end{equation}
where $\bar{q}_j(t)=\int {\rm d}\tilde{\omega} 
{\rm e}^{{\rm i}\tilde{\omega} t} q_j(\tilde{\omega})/2\pi$.
Following~\cite{Dyson53}, this problem reduces to the 
diagonalization of an
antisymmetric matrix. In what follows we review the fundamental steps. We
define the variables $s_1,\ldots,s_{N-1}$, such that
$\dot{s}_j=\sqrt{\Lambda_j} (\bar{q}_{j+1} - \bar{q}_{j})$. After 
substituting into Eq.~(\ref{Eq:2a}) and integrating, we get the first-order
differential equation $\dot{\bar{q}_j}=\sqrt{\Lambda_{j}}s_j
-\sqrt{\Lambda_{j-1}}s_{j-1}$. We introduce the variables
$u_1,\ldots,u_{2N-1}$, such that $u_{2j-1}=\bar{q}_j$, $u_{2j}=s_j$, and
the new matrix elements $\tilde{\Lambda}_j$, such that
$\tilde{\Lambda}_{2j}=\tilde{\Lambda}_{2j-1}=\Lambda_j$. With
these definition, we have now a set of $2N-1$ first-order
differential equations
$$
\dot{u_j}=\sqrt{\tilde{\Lambda}_j} u_{j+1}-\sqrt{\tilde{\Lambda}_{j-1}} u_{j-1}
$$
which are now defined by the $(2N-1)\times(2N-1)$
antisymmetric matrix $\bar{\Lambda}$, whose elements
are $\bar{\Lambda}_{j+1,j}= -\bar{\Lambda}_{j,j+1}
={\rm i}\sqrt{\tilde{\Lambda}_j}$.
The characteristic frequencies $\tilde{\omega}_j$ of the chain are
the square roots of the characteristic roots of $-\Lambda$,
of which one is 0 and the other $2N-2$ come in pairs 
$\pm \tilde{\omega}_j$. \\

Using the matrix $\bar{\Lambda}$, in the limit $N\gg 1$ the spectrum of
characteristic frequencies $M(\tilde{\omega}^2)$, that is defined as the
proportion of roots with $\tilde{\omega}_j^2<\tilde{\omega}^2$, 
and the corresponding density
$D(\tilde{\omega}^2)=dM(\mu)/d\mu|_{\mu=\tilde{\omega}^2}$, 
can be derived from the characteristic function $\Omega(\theta)$, which is
defined as~\cite{Dyson53}
\begin{eqnarray}
\label{Char:1}
\Omega(\theta)
&=&\lim_{N\to\infty}\frac{1}{2N-1}\sum_{j=1}^{2N-1}
\log(1+\theta\tilde{\omega}_j^2)
\end{eqnarray}
where $\theta$ is a complex variable. The density of characteristic
frequencies $D(\mu)$ and the spectrum $M(\mu)$ are found by using the
properties of the analytic continuation of $\Omega(\theta)$,
\begin{eqnarray*}
{\rm Re}\left[\frac{1}{{\rm i}\pi}\lim_{\epsilon\to 0}\Omega(-z+{\rm
      i}\epsilon)\right] 
=\int_{1/z}^{\infty}{\rm d}\mu D(\mu)=1-M(1/z)
\end{eqnarray*}
from which Eq.~(\ref{D1z}) is obtained.

\section{An ansatz for the secular equation: Slowly varying envelope}

\noindent In this appendix we discuss an ansatz based on the phononic
solution for a translationally invariant crystal, where we assume that the
envelope, superimposed on the amplitude and the phase of the phononic solution, 
varies slowly as a function of the
position. This ansatz is supposed to be valid for long--wavelength excitations
and chains with $N\gg 1$ atoms.

The exact solution of~(\ref{Eq:ax}-\ref{Eq:z}) in the limit of uniform
charge distribution with constant spacing $a$, 
$K_{i,j}/m=\kappa/|i-j|^3$ with $\kappa=2Q^2/a^3$, is~\cite{Ashcroft}  
\begin{equation}
\label{Mode:0}
q_j\sim {\rm e}^{{\rm i}(kja-\omega t)}
\end{equation}
Here, $\omega$ is the eigenfrequency, $k$ the wavevector, and the
boundary conditions for the uniform chain 
are fixed so that $q_N=q_0$, giving
${\rm e}^{{\rm i}kNa}=1$. Hence, $ka=2\pi l/N$ 
with $l=0,1,\ldots,N-1$. The dispersion relation is
obtained by substituting (\ref{Mode:0}) into (\ref{Eq:ax}) with
$K_{i,j}/m=\kappa/|i-j|^3$. If the interaction is determined by the 
nearest-neighbours, while the interaction with the other ions is neglected,
then the eigenfrequencies are given by 
\begin{equation}
\label{Dispersion:0}
\sqrt{\omega^2-\nu^2}=2\sqrt{\kappa}\left|\sin\frac{ka}{2}\right|
\end{equation}
Note that $\omega$ depends on $ka$, which takes the values $ka=2\pi l/N$. \\

We now construct from this solution an ansatz for 
the inhomogeneous chain, in the limit of slowly-varying spring constant
$K_{j,j+1}$ over the wavelength of the propagating perturbation. We assume
thus the local solution:
\begin{equation}
\label{Ansatz}
q_j=A_j{\rm e}^{{\rm i}(kja-\omega t)}
\end{equation}
where $ka$ is a constant, i.e. we have assumed $k_jx_j^{(0)}=jka$. The
non-linear variation of the phase with the site is included in $A_j$.
The ansatz~(\ref{Ansatz}) assumes that it is possible to write the
solution as the product of a slowly varying amplitude and phase, represented
by the factor $A_j$, and a fast oscillating part, that is the exponential term
in~(\ref{Mode:0}). The validity of this assumption is checked
later on.

By substituting (\ref{Ansatz}) into (\ref{Eq:ax}), keeping only the nearest
neighbour interaction, we obtain:
\begin{eqnarray}
\label{EqAns:1}
(\omega^2-\nu^2)A_j
&=&\kappa_j
\left(A_j-A_{j+1}{\rm e}^{{\rm i}ka}\right)\\
&+&
\kappa_{j-1}\left(A_j-A_{j-1}{\rm e}^{-{\rm i}ka}\right)
\nonumber
\end{eqnarray}
where $\kappa_j=K_{j,j+1}/m$. We make use of the fact that $\kappa_j$ varies slowly
with the position, which allows to expand $A_j$ and $\kappa_j$ in $\delta j$,
of order unity, according to
\begin{eqnarray}
A_{j\pm 1}
&\approx& A \pm \delta A
\label{A:approx}\\
\kappa_j
&   =   &\kappa+\Delta \kappa+\delta\kappa_j/2\\
\kappa_{j-1}
&   =   &\kappa+\Delta \kappa-\delta\kappa_j/2
\label{kappa:0}
\end{eqnarray}
where $\delta A= \frac{\partial A}{\partial j}\delta j$
with $\delta j=\pm 1$. The spring constant has been
divided into three contributions: $\kappa$ is constant and fulfills
(\ref{Dispersion:0}), $\Delta\kappa$ is also constant and
$\delta \kappa_j=\kappa^{\prime}\delta j$, with $\kappa^{\prime}=\frac{\partial
  \kappa_j}{\partial j}$, such that $\kappa+\Delta \kappa\gg \delta
\kappa_j$. The chosen variation of $\kappa_j$ reflects the symmetry under
reflection of the crystal, such that the ion at the center is characterized by
equal couplings on both sides (which implies the condition $\delta\kappa=0$).
The term $\Delta \kappa$ does not depend on $j$, and it is the
correction to $\kappa$, the spring constant fulfilling the dispersion relation for
the case of a uniform chain: $\Delta \kappa\approx \kappa_j-\kappa$.
Substituting (\ref{A:approx}-\ref{kappa:0}) into (\ref{EqAns:1})
yields 
\begin{eqnarray*}
4\Delta\kappa A \sin^2\frac{ka}{2}
-2{\rm i}\kappa \delta A\sin ka-2{\rm i}A\delta\kappa\sin ka=0
\end{eqnarray*}
where we have applied the dispersion relation (\ref{Dispersion:0}).
Since there is only one zero-order term in $j$, it is evident that
$\Delta\kappa=0$: Thus, there is no zero-order correction to the spring
constant $\kappa$ of the uniform chain, and its value in this order is
accounted for by~(\ref{Dispersion:0}). The first order
expansion gives thus a differential equation (valid for $\sin ka\neq 0$)
\begin{equation}
\frac{\delta A}{A}=-\frac{\delta \kappa}{2\kappa}
\end{equation}
which admits the solution
\begin{equation}
\label{Sol:1}
A_j=A_0\sqrt{\frac{\kappa_0}{\kappa_j}}
\end{equation}
The eigenmode has then the form:
\begin{equation}
\label{Sol:2}
q_j=\sqrt{\frac{\kappa_0}{\kappa_j}}{\rm e}^{{\rm i}kja-{\rm i}\omega t}
\end{equation}
and the wave vector takes the same values as for a uniform crystal with
half-periodic boundary conditions, taking into account the symmetry
under reflection,
\begin{equation}
\label{Boundary}
kNa =\pi l
\end{equation}
where $k=k(\omega)$ is given by Eq~(\ref{Dispersion:0}),
and ${l=0,1,\ldots,N-1}$. Therefore, this ansatz gives simply
a variation of the amplitude of the displacement of the ion as
a function of the local spring constant, but no change in the
phase, which is the same as the one of a uniform crystal. The
amplitude is smaller at the center of the chain, where
the ions are closer and the coupling constant is larger.
Taking $\kappa_0=\Lambda_0$ as given by~(\ref{lambda:0}), i.e.\ the maximum value
of the spring constant in the chain, and using~(\ref{Dispersion:0}),
we obtain the maximal frequency as given in~(\ref{w:max}). A similar argument
can be developed for the transverse frequency.

\begin{center}
\begin{figure}[h]
\epsfxsize=0.3\textwidth
\epsffile{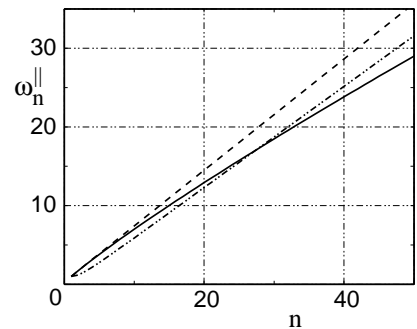}
\caption{Long wavelength part of the axial eigenmodes spectrum for $N=1000$
  ions. The dot-dashed
  line gives the spectrum evaluated using~(\ref{Sol:2}-\ref{Boundary}), the
  solid line the numerical results and the dashed line the spectrum evaluated
  using Eq.~(\ref{omega:ax}).}
\label{Fig:Last}
\end{figure}
\end{center}

Figure~\ref{Fig:Last} compares the prediction of the slowly--varying 
ansatz with the numerical results and the Jacobi polynomials results given by
Eq.~(\ref{omega:ax}). Here, it is obvious that the Jacobi polynomials
provide a better approximation for the spectrum at the lowest eigenfrequencies. 
The curve evaluated from Eq.~(\ref{Dispersion:0}) using Eqs.~(\ref{Ansatz})
and~(\ref{Boundary}) 
lies close to the spectrum evaluated numerically for a larger range of
modes, but it does not reproduce its asymptotic behaviour for $N\to\infty$. 

\end{appendix}

\end{document}